\documentclass[12pt]{iopart}

\usepackage{graphicx}
\usepackage{dcolumn}
\usepackage{bm}

\def\be{\begin{equation}}
\def\ee{\end{equation}}
\def\ben{\begin{displaymath}}
\def\een{\end{displaymath}}
\def\ba{\begin{eqnarray}}
\def\ea{\end{eqnarray}}
\def\nn{\nonumber}
\def\os{Ozsv\'{a}th and Sch\"ucking }
\def\oss{Ozsv\'{a}th and Sch\"ucking's }

\begin{document}
                         

\title[G\"odel Light Cone]{Revisiting the Light Cone of the G\"odel Universe}
\author{G.~Dautcourt}
\address{Max Planck Institut f\"{u}r Gravitationsphysik,
Albert-Einstein-Institut,\\ 
Am M\"{u}hlenberg 1, D-14476 Golm, Germany\\
E-mail: daut@aei.mpg.de} 
\author{M.~Abdel-Megied} 
\address{Mathematics Department, Faculty of Science, Minia University,\\   
El-Minia, Egypt\\ 
E-mail: amegied@mailer.eun.eg}
\date{\today}

\begin{abstract} 
The structure of a light cone in the G\"odel universe is studied.   
We derive the intrinsic cone metric,  
calculate the rotation coefficients of the ray congruence
forming the cone, determine local differential 
invariants up to second order, describe the crossover
(keel) singularities  
and give a first discussion of its focal points. 
Contrary to many rotation coefficients, some inner differential 
invariants attain simple finite standard values at focal singularities.
\end{abstract}


\section{Introduction}
G\"odel's rotating cosmological model \cite{godel49}, \cite{godel52} 
is one of the most interesting solutions of Einstein's field equations 
with negative $\Lambda$-constant, 
particularly in view of its contribution to our understanding of
rotation in relativity and its signs of causality breakdown due to the 
existence of closed timelike curves 
\cite{hawell73}, \cite{rooman98}, 
\cite{os03}.

The G\"odel solution is also of interest
for a study of light ray caustics, which are basic for a discussion
of the strong lensing effects in the Universe \cite{schn92},\cite{perlick04}.
There are now many papers discussing singularities on 
characteristic manifolds 
of the Einstein field equations, 
mainly based 
on powerful mathematical theorems of Lagrangian and Legendrian maps
\cite{cork83},\cite{friedr83},\cite{arnold90}, \cite{ehl00}, \cite{fritt03}.
One may also mention an older article by Laurent, Rosquist and Sviestins
\cite{laur81}, where the cone of an Ozsv\'{a}th class III metric \cite{os70} was
studied. This metric already includes the G\"odel metric as a particular case.

Focal subsets  (caustics)
are likely to be present on null hypersurfaces, if a weak energy 
condition holds for its lightlike generators (see \cite{pen65}, 
\cite{daut65}), hence almost always in realistic astrophysical
or cosmological situations. 
They have often a      
complicated structure and are an obstacle for attempts to solve the 
characteristic
initial value problem for Einstein's field equations  
numerically, since integration along null geodesics runs into difficulties at
caustics \cite{cork83},
\cite{friedr83},\cite{winic98}. It would be extremely helpful if
existent algorithms could be modified or replaced 
to allow numerical processing through 
such singularities. 
A preliminary step is 
to study caustics in exact solutions of the field equations.

In this connection the G\"odel light cone could be a useful object, 
since here the behaviour of light rays 
is already sufficiently complex 
to give an impression of features which we can expect 
in more realistic geometries,   
and on the other side it is simple enough to allow a
complete analytical treatment.
Due to the five-dimensional group of isometries admitted by the G\"odel 
metric, which has a four-dimensional transitive subgroup, 
all light cones have the same internal structure.
 
A first discussion of the inner geometry of the G\"odel cone was 
given by us    
in 1972 \cite{abdel72a}, \cite{abdel72b}, based on the integration of 
geodesics performed by Kundt in 1956 \cite{kundt56}. 
Use of computer based formula manipulation technique has shown, that 
a number of complicated relations can be simplified considerably.
In particular, the structure of caustics and the resulting startling 
cyclic lens effects found in
\cite{abdel72a} and \cite{abdel72b} now became more transparent. 
The lens effects arise from a quasi-periodic re-focussing of the generators
and are surprisingly similar to those discussed by \os for the light cone 
of a plane gravitational wave propagating in vacuum, one 
of their anti-Mach metrics \cite{os62}.

We apply the geometrical decription of null hypersurfaces developped in 
\cite{daut65},\cite{daut67}. After integrating the null geodesics in 
Section II, we derive the intrinsic metric of the G\"odel cone in 
Section III, 
calculate and discuss its rotation coefficients and differential 
invariants in Section IV and turn to a description of caustics      
in Section V. 
While rotation coefficients of the light ray 
congruence forming the cone have as a rule singularities on     
focal surfaces or  keel points, some local inner differential 
invariants have simple {\it finite} limits there. It is 
interesting that this feature - with the same asymptotic values 
of invariants at singularitites - 
has shown up in all nontrivial light cones 
studied so far by us. 
A method to relate intrinsic
cone coordinates to the angles $(\theta,\phi)$ on the observer 
sky is described in an Appendix.

\section{Light rays in the G\"odel universe}
\subsection{General Congruence}

G\"odel's stationary solution of Einstein's field equations 
with cosmological constant 
describes the gravitational 
field of a uniform distribution of rotating dust matter, where 
- loosely speaking - the gravitational attraction of matter and the added 
attractice force of a negative $\Lambda$-constant is compensated by the 
centrifugal force of rotation.
Hawking and Ellis 
\cite{hawell73} introduce the G\"odel metric with the coordinates
$t,x,\bar{y},z$ as (we have exchanged $x,y$ to reach conformity with
our notation, the signature will be taken as (-1,1,1,1), 
and the conventions of the Misner-Thorne-Wheeler book will be adopted): 
\ben
ds^2= -dt^2+d\bar{y}^2- \frac{1}{2}e^{2\bar{y}/b}dx^2 +dz^2 
-2e^{\bar{y}/b}dtdx. 
\een
The matter density is given by 
$\kappa\mu = -2\Lambda = 1/b^2$.
With $y= \sqrt{2}b e^{-\bar{y}/b}$ 
one obtains the form of the metric used in \cite{abdel72b} and employed 
also here:
\be
ds^2= -(dt+\frac{\sqrt{2}b}{y}dx)^2 +\frac{b^2}{y^2}(dx^2+dy^2) +dz^2
\ee
As first shown by Kundt \cite{kundt56}, the differential equations 
$x''^{\mu}+ \Gamma^\mu_{\rho\sigma}x^\rho{'} x^\sigma{'} = 0$ 
for the geodesics admit the first integrals
\ba
t' \equiv  x^{(0)}{'} &=& (-c_2/\sqrt{2}+\sqrt{2}y)/c_0,  \label{te} \\
x' \equiv  x^{(1)}{'} &=& y (c_2- y)/(bc_0), \label{xe}  \\
y' \equiv  x^{(2)}{'} &=&  y (x -c_1)/(b c_0), \label{ye}  \\
z' \equiv  x^{(3)}{'} &=&  c_3/c_0, \label{ze}
\ea
Here the prime denotes the derivative with respect to a running parameter
$s$ on the geodesic, i.e., $s$ is the proper time or invariant length for a 
non-null geodesic and an affine parameter for light rays. The 
integrals depend on four parameters $c_0,c_1,c_2,c_3$ and are subject 
to the normalization condition
$g_{\mu\nu}x^{\mu}{'}x^{\nu}{'}= constant$, 
where the constant is -1 for a timelike, 1 for a spacelike and 0 for a null 
geodesic.
For the null geodesics discussed here, different values of $c_0$ correspond 
only to different definitions of the affine parameter, so we assume  
$c_0=1$  subsequently.
With (\ref{te})-(\ref{ze}), the normalization condition 
becomes for null geodesics
\be
(x-c_1)^2+(y-c_2)^2 = \frac{1}{2}c_2^2- c_3^2 \equiv c_4^2, \label{null}
\ee 
thus the projection of null geodesics into the $xy$-plane are confined to a 
circle with radius $|c_4|$. Light can move to arbitrary large distances only
in the $z$-direction, the direction of the rotation axis.  
Equation (\ref{null}) is solved with 
\be
 x = c_1+c_4 \sin{\Phi}, ~ y = c_2+c_4\cos{\Phi}, \label{xy}
\ee
where $\Phi(s)$ is an unknown function. 
Since $y \geq 0$, no circle point can lie below the $y$-axis. 
The parameter range for null geodesics is thus constrained by 
\be c_2 \geq \sqrt{2}|c_4|, \label{ineq1}
\ee
corresponding to points below the line $ACD$ in the upper half
plane and above the line $AEF$ in the lower half plane 
of Fig. 1.

(\ref{xe}) and (\ref{ye}) lead 
to a single differential equation for $\Phi:$
\be
b\Phi' +c_4 \cos{\Phi} +c_2 = 0. \label{deqn}
\ee
We first shortly discuss the particular case $c_4 =0$, when the 
circle in the $xy$-plane shrinks to a point. Here (\ref{te})-(\ref{ze})
have the solution 
\be
t = \frac{c_2}{\sqrt{2}}s+t_0,~x=c_1,~y=c_2,~z=\frac{c_2}{\sqrt{2}}s+z_0, 
\label{rr} 
\ee
thus through every point $P$ of the G\"odel universe passes one 
exceptional light ray: It is the light ray sent or received 
by a comoving observer at $P$ in or opposite to the direction 
of the local rotation axis.   
Returning to the general case, integration of (\ref{deqn}) gives with an 
integration constant $k_1$
\be
\tan{\frac{\Phi}{2}}=  -\sqrt{\frac{c_2+c_4}{c_2-c_4}} 
   \tan{(\sqrt{c_2^2-c_4^2}\frac{s+k_1}{2 b})}. \label{fexpr}
\ee
The inequality (\ref{ineq1}) ensures that the roots are real.
To simplify the representation, we define a new real constant $k$
\be
    k = \frac{k_1}{2b}\sqrt{c_2^2-c_4^2},
\ee
and introduce a new affine parameter $w$ instead of $s$:
\be
    w = \frac{s}{2b}\sqrt{c_2^2-c_4^2}.
\ee
Integrating also the remaining equations (\ref{te}),(\ref{ze}), 
one finally obtains ($\epsilon=\pm 1$, since $c_3$ can have both
signs)
\ba
t(w) &=& 2\sqrt{2}b \arctan{(\sqrt{\frac{c_2+c_4}{c_2-c_4}}\tan{(w+k)})}
 -\frac{\sqrt{2}bc_2w}{\sqrt{c_2^2-c_4^2}} + c_5, \label{gt} \\
x(w) &=& c_1 -\frac{c_4\sqrt{c_2^2-c_4^2} 
\sin{(2w+2k)}}{c_2-c_4\cos{(2w+2k)}}, \\
\label{gx}
y(w) &=& \frac{c_2^2-c_4^2}{c_2-c_4\cos{(2w+2k)}}, \\ \label{gy}
z(w) &=& \sqrt{2}\epsilon bw\frac{\sqrt{c_2^2-2c_4^2}}{\sqrt{c_2^2-c_4^2}}
+c_6 \label{gz}
\ea  
as parameter representation of the null geodesics. 
Counting the number of independent parameter one sees that 
(\ref{gt})-(\ref{gz}) is the generic 
null congruence of the G\"odel cosmos. Its explicit form helps to            
answer questions on null geodesics in the G\"odel cosmos.
For example, one can easily conclude that there are 
no closed null geodesics, which would require
$x^\mu(w)=x^\mu(w_1)$
for some values $w$ and $w_1$: Taking first $\mu =3$, 
Equation (\ref{gz}) shows that 
$c_3=0$ or $c_2=\sqrt{2}c_4$ is needed, which corresponds to $z=const$.  
An inspection of the relation for $\mu =1,2$ shows, that these
relations can be satisfied by means of periodic functions.  
Thus a subset of null geodesics may return to the same
space point, but the point is (repeatedly) reached at 
{\it different} times $t$, due to the aperiodic term proportional 
to $w$ in (\ref{gt}).

\subsection{Light cone geodesics}
We are here interested in those null geodesics, which form a cone 
with the 
vertex at a point $P_0$ with the coordinates $ (0,0,b,0)$, say. $P_0$          
corresponds to the origin of the Hawking-Ellis coordinates. Since all 
light cones of the G\"odel cosmos have the same intrinsic structure, 
we could have chosen any other origin in principle. Furthermore, for 
definiteness, the past cone ($w>0$) will be considered.  
Assuming $w=0$ at the vertex, we have four relations which will be 
used to determine $c_1,c_5, k$ and $c_6$ in terms of the remaining 
parameters $c_2$ and $c_4$:
\ba
c_1 &=& \frac{c_4\sqrt{c_2^2-c_4^2}\sin{2k}}{c_2-c_4\cos{2k}}, 
\label{c1} \\
c_5 &=& -2b\sqrt{2} \arctan{(\sqrt{\frac{c_2+c_4}{c_2-c_4}}\tan{k}}),
\label{c5} \\
\cos{2k} &=& \frac{c_4^2-c_2^2+bc_2}{bc_4}, \label{k} \\
c_6 &=&0.\label{c6}
\ea
The requirement that the $c_1,c_5,k$ and $c_6$ exist and are real restricts 
$c_2$ and $c_4$ beyond (\ref{ineq1}). 
In \cite{abdel72a},\cite{abdel72b} a pair $(u,v)$ of transversal
parameters was introduced to replace $(c_2,c_4)$: 
\be
u^2=\frac{c_4+c_2-b}{c_4-c_2+b},~v^2=\frac{c_2+c_4}{c_2-c_4}.\label{uvdef}
\ee
Inverting, we have
\ben
c_2= \frac{b(1+u^2)(v^2+1)}{2(u^2+v^2)}, 
~c_4= \frac{b(1+u^2)(v^2-1)}{2(u^2+v^2)}.\nn 
\een 

The map $(c_2,c_4)  \rightarrow  (u,v)$ is not everywhere 
regular, since the functional determinant
$
\frac{\partial c_2}{\partial u}\frac{\partial c_4}{\partial v}
-\frac{\partial c_2}{\partial v}\frac{\partial c_4}{\partial u} 
= -2uvb^2\frac{(u^2+1)(v^2-1)}{u^2 +v^2}$
has for real $u,v$ zeros at $u=0, v=0,v^2=1$. The first two arise from using
squares on the lhs of (\ref{uvdef}), the singularity $v^2=1$ or $c_4=0$ 
corresponds to the exceptional ray introduced above.
In terms of $u$ and $v$, (\ref{k}) can be written
$\cos{2k}= \frac{v^2-u^2}{v^2+u^2}$.
It is seen from this equation, that $u^2$ cannot be negative for geodesics 
forming the $P_0$-cone, this also applies to $v^2$ because
of (\ref{ineq1}). Thus $u$ and $v$ as introduced by (\ref{uvdef}) are 
real. 

For more information we refer to Fig. 1.
\begin{figure}[tbp]
\includegraphics[width=5.0cm]{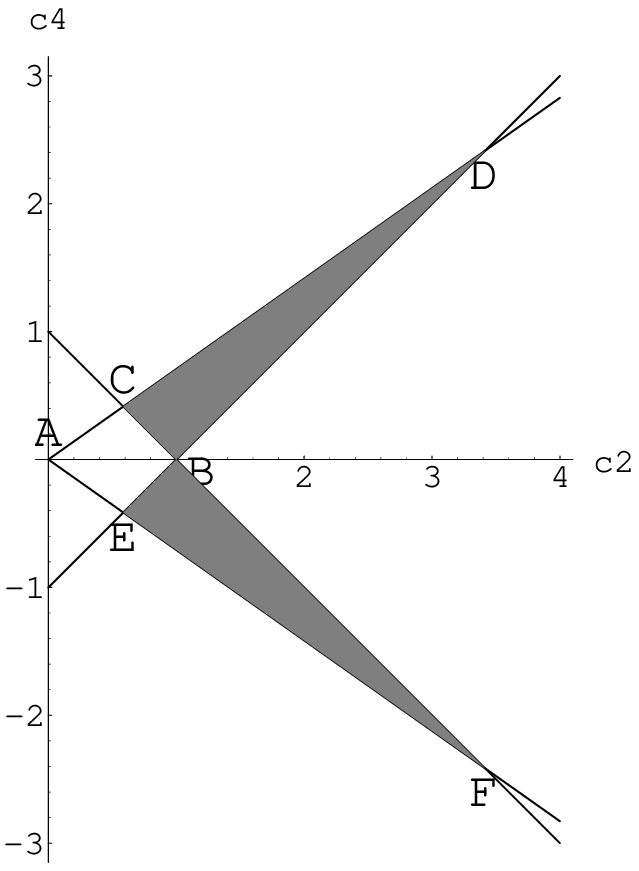}
\caption[fig1]\
\small \baselineskip=11pt
{ $c_2$-$c_4$-parameter plane for null geodesics in the G\"odel universe 
($b=1$ assumed). Null geodesics  have their 
integration constants $c_2$ and $c_4$ between the lines $ACD$ and $AEF$. 
Those forming the (past) light cone through $P_0$ are confined to the 
shadowed triangles, with $c_4>0$ for the western hemisphere and  
$c_4<0$ for the eastern hemisphere of the observer sky. 
But only 
the northern hemisphere is mapped 1:1 to the two 
triangle regions, for the missing southern hemisphere a second copy of 
the figure is needed.
To obtain the topology of a sphere, certain boundary lines of the 
triangles in both copies must be identified pointwise. 
The lines $CD$ and $EF$ correspond to parts of the equator, 
the pole ray at $B$ is the exceptional ray towards the rotation direction.}

\hrulefill
\end{figure}
In this parameter plane the points $A$ through $F$ correspond to   
coordinate pairs $(c_2,c_4)$ given by
\ba
\fl \qquad &&A = (0,0),~~ B = b(1,0),~~
C = b(\frac{\sqrt{2}}{\sqrt{2}+1},\frac{1}{\sqrt{2}+1}),~~      
D = b(\frac{\sqrt{2}}{\sqrt{2}-1},\frac{1}{\sqrt{2}-1}), \nn \\
\fl \qquad &&E = b(\frac{\sqrt{2}+2}{\sqrt{2}+1},-\frac{1}{\sqrt{2}+1}),~~      
F = b(\frac{\sqrt{2}}{\sqrt{2}-1},-\frac{1}{\sqrt{2}-1}). \nn
\ea
Allowed parameters $c_2,c_4$ for generators are subject to (\ref{ineq1}), 
and lie   
above the line $AEF$ as well as below the line $ACD$.
$v = const$ is the equation of straight lines ("parallels") 
starting at $A$, with $v$ ranging from $v=1$ (the $c_2$-axis) 
to $v=1+\sqrt{2}$ (the line $ACD$) in the upper half plane.
For $v<1$ the lines $v=const$ lie in 
the lower half plane  $c_4 < 0$,
ranging from $v=1$ through $v=1/(1+\sqrt{2})$ (the line $AEF$).
Curves with $u=const$ are straight lines ("meridians") through the 
point $B$, ranging from
$u=0$ ($BC $) to $u \rightarrow \infty $ ($BD$) in the upper half 
plane. We shall find it appropriate (Appendix B) to take $u$ negative in one  
hemisphere. 
As discussed subsequently, the cone generators 
cover the 
two shadowed triangles in Fig~1, which correspond to certain quadrants
on the observer sky, 
e.g., the lines $EF$  and $CD$ form part of the equator.
Two copies of the figure are required to cover the full observer 
sphere. For details we refer to Appendix B.

Returning to the cone representation, 
the parameters $c_5,c_1,k$ and $c_6$ 
can be written in terms of $u$ and $v$ in a compact form as solutions of 
(\ref{c1})-(\ref{c6}):
\ba
c_1 &=& -b\frac{u(1-v^2)}{u^2+v^2},\nn  \\
c_5 &=& -2\sqrt{2}b\arctan{u}, \nn \\
\tan{k} &=& \frac{u}{v}, \nn \\
c_6 &=& 0.\nn 
\ea
Substituting these values into (\ref{te})-(\ref{ze}), 
one obtains as a parameter representation of the light cone 
through $P_0$:           
\ba
\fl \qquad t&=& t(u,v,w) = -\frac{bw}{\sqrt{2}}(v+\frac{1}{v})        
  +2\sqrt{2}b\arctan{\frac{(u^2+v^2)\tan w}{v(u^2+1) +u(v^2-1)\tan(w)}},
 \label{tc} \\
\fl \qquad x&=&x(u,v,w) =        
  b(v^2-1)\frac{\sin{w}(v(u^2-1)\cos{w} +u(v^2+1)\sin{w})}
{(v\cos{w} -u\sin{w})^2 +v^2(u\cos{w}+v\sin{w})^2},  \label{xc} \\
\fl \qquad y&=&y(u,v,w) =        
  \frac{bv^2(u^2+1)}{(v\cos w-u\sin w)^2+v^2(u\cos w+v\sin w)^2},
\label{yc} \\
\fl \qquad z&=&z(u,v,w) =  
\frac{b\epsilon w}{\sqrt{2}v}\sqrt{6v^2-1-v^4}.   \label{zc}
\ea
Positive values of the affine parameter $w$ correspond to the past
light cone, negative values to the future cone. The sign $\epsilon$
distinguishes between the northern ($\epsilon =1$) 
and southern ($\epsilon = -1$) hemisphere of the observer sky. 
We note the following invariance property of the system (\ref{tc})-(\ref{zc}):
Substituting $-1/u$ for $u$ and $1/v$
for $v$ leads to the same geodesics:
\be
x^\mu(u,v,w) = x^\mu(-\frac{1}{u},\frac{1}{v},w) \label{map}
\ee
The same map 
sends also $(c_2,c_4)$ into $(c_2,-c_4)$. Some geodesics of the
parallel $v=1.5$ are plotted in Fig. 2.

\begin{figure}[tbp]
\includegraphics[width= 6.0cm]{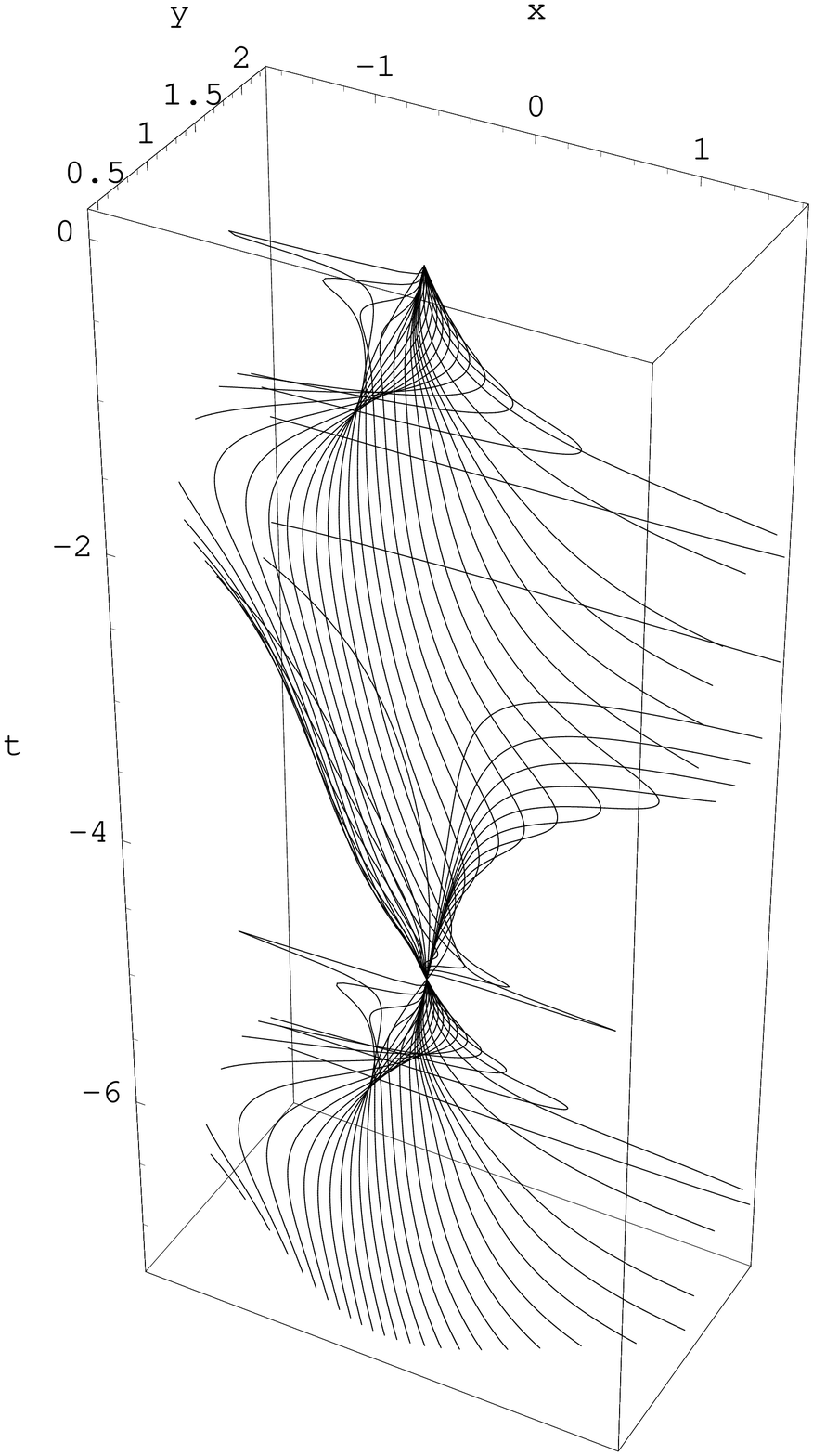}
\caption[fig2]\
\small
\baselineskip=10pt
\smallskip
{
The twisted generators of the past light cone through $P_0$ are shown in a 
fictitious 
Euklidean 3-space ${\cal R}^3$. Plotted are the coordinates 
$x(u,v,w),y(u,v,w),t(u,v,w)$ from (\ref{tc})-(\ref{yc}) for 
rays on the parallel $v=1.5$ and several $u$ with constant 
separation $\pi/30$ in the meridian angle $\phi =2 \arctan{u}$ 
corresponding to $u$. The $z$-coordinate is suppressed, 
keels appear as points, focal surfaces as lines. 
The affine parameter ranges from 
$w=0$ at the vertex (top) to $w=1.5 \pi$ (bottom), the plot ends before the 
second keel line at $w=2\pi$ is reached.-  Distances cannot be represented 
correctly in such a projection, the type of caustics is preserved however, 
since the map from the curved $V^3$ to ${\cal R}^3$ is a diffeomorphism. }
 
\hrulefill
\end{figure}

Knowing the tangential vector $\frac{dx^\mu}{dw}$ of the past light cone
from (\ref{tc})-(\ref{zc}) or from (\ref{te})-(\ref{ze}), 
one can calculate the redshift $z$ of distant objects from the 
well-known relation 
\ben
1+z=\frac{ (\frac{dx^\mu}{dw}V^\nu g_{\mu \nu})_{emitter}}
         { (\frac{dx^\mu}{dw}V^\nu g_{\mu \nu})_{observer}}. 
\een
With $V^\mu= \delta^\mu_0$ and (\ref{tc})-(\ref{zc}) one finds $z=0$:
As noted already by G\"odel in his original paper \cite{godel49}, distant 
objects comoving with the cosmic fluid would show no redshift, 
proving that this model cannot represent the real Universe.
                                                 
\section{Light cone metric} 
The spacetime coordinates of light-rays through $P_0$, (\ref{tc})-(\ref{zc}),
 depend on the affine 
parameter $w$ as well as on two quantities $u,v$. While $w$        
determines a position on a light ray, $u$ and $v$ label a  ray.
The tripel $w^i \equiv (w,u,v)$  may therefore be used as intrinsic 
coordinates on the cone. Below we will see how $u$ and $v$ 
are related to the angles $\theta, \phi$ on the sky of a comoving 
observer at $P_0$, who wants to fix an event on his past light cone.
The intrinsic three-dimensional metric of the light cone at $P_0$ can be 
found from (\ref{tc})-(\ref{zc}) by means of  
\ben
\gamma_{ik} = g_{\mu\nu}\frac{\partial x^\mu}{\partial w^i}
\frac{\partial x^\mu}{\partial w^k},
\een
where $w^i=(w,v,u)$. Since $\frac{\partial x^\mu}{\partial w}$ is the 
tangential vector to the cone and hence null, the components of the
cone metric can be reduced to a two-dimensional metric $\gamma_{AB}$:    
\ba
\gamma_{00} &=& 0,   \label{im000}\\
\gamma_{0A} &=& 0,    \label{im00A}\\  
\gamma_{22} &=& \frac{b^2}{v^4f_2}(w^2f_1^4+2wf_2f_1^2\sin{ w }\cos{ w }      
   +f_2^2\sin^2{ w }+f_2f_1^2\sin^4{ w }),  \label{im022}\\
\gamma_{23} &=& \frac{b^2f_1^3\sin^2{w}(w-\sin{ w }\cos{ w })}{v^4(u^2+1)},
 \label{im023}\\
\gamma_{33} &=& \frac{b^2f_1^2\sin^2{w}(4v^2-f_1^2\sin^2{ w })}{v^4(u^2+1)^2},
  \label{im033}\\ 
\ea
where we have introduced the two functions
\be
f_1(v) = v^2-1,~ f_2(v) = 6v^2 -1-v^4, 
\ee
which are useful to compactify expressions.

In general, the $\gamma_{AB}$ form an one-dimensional sequence of 
positive-definite two-dimensional metrics on the cone, parametrized
by the affine parameter $w$. They represent 
metric spheres in the neighbourhood of the vertex $w=0$, but become 
progressively deformed for increasing parameters $w$. 
On some 
subsets of the cone the inner metric degenerates additionally, 
i.e. the two-dimensional determinant $\gamma$ becomes zero. This 
signifies an intersection of light rays forming the cone, if the 
zero of $\gamma$ is not caused by a coordinate singularity. In focal 
points or caustics geodesics with infinitesimally differing values 
of $u,v$ meet, while in crossover or keel points (we have taken the 
latter notation
from a paper by  M.~Riesz \cite{riesz}) 
the intersecting geodesics 
may have quite different transversal parameters.
Sets of focal points form in general two-dimensional focal surfaces on 
a null hypersurface. 
For the G\"odel cone the determinant $\gamma$ can be represented in 
a compact form:
\ba
  \gamma &\equiv& |\gamma_{AB}| = h^2, \\
  h &\equiv& \frac{2b^2f_1p(v,w)\sin{w}}{(u^2+1)\sqrt{f_2}v^3}, \\
  p(v,w) &\equiv& f_1^2w\cos{w} + f_2\sin{w} \label{pfunc} 
\ea
(abbreviations are deliberately chosen in this paper 
to
compactify expressions).
The simple expression for $\gamma$ 
allows to pick up the cone singularities easily. 
Apart from the 
vertex $w=0$, $\gamma$ vanishes periodically for $w=n\pi$, $n$ an 
integer, which is  
similar to the behaviour of light cones in a closed Robertson-Walker model
with the time extended to several cycles. 
But contrary to the RW case,  the two-dimensional spacelike 
surfaces $w=const$ do not shrink
to points at $w=n\pi,~n\neq 0$, but rather to spacelike lines, the keel lines,
discussed in Section V. Further zeros of $\gamma$ are given by
$p(v,w)=0$, which is the equation of the focal surfaces, also discussed in 
Section V. The remaining zeros of $\gamma$ are given by $f_1=0$ or $v=1$,
corresponding to the exceptional pole rays, and $u \rightarrow \infty$. The 
latter is related to a coordinate singularity. 

For later use it is appropriate to introduce a second function 
$q(v,w)$, with the property of being not negative: 
\be
 q(v,w)\equiv 4(1+f_1)-f_1^2\sin^2{w}. \label{qfunc}
\ee
$q$ vanishes if and only if $f_2/f^2_1 = -\cos^2{w}$. Since the lhs is not
negative, the condition $q=0$ holds only for $\cos{w}=0$ (or $w=(1+2n)\pi/2$,
$n$ integer), and for $f_2=0$, or $v=1+\sqrt{2}$. Since at these points also
$p=0$, the equation $q=0$ represents curves on the light cone, where the 
equator  rays $v=1+\sqrt{2}$ meet the $n$th focal surface (Section IV). 
With $q$,  the intrinsic cone metric can also be written as                          
\ba
\gamma_{22} &=& \frac{b^2}{v^4f_2\cos^2{w}}(p^2 -2pf_2\sin^3{w}
+qf_2\sin^4{w}), \label{im22} \\
\gamma_{23} &=& \frac{b^2f_1\sin^2{w}}{v^4(u^2+1)\cos{w}}(p
-q\sin{w}), \label{im23}\\
\gamma_{33} &=& \frac{b^2f_1^2\sin^2{w}}{v^4(u^2+1)^2}q,\label{im33}
\ea 
allowing to check easily that the rank of 
$\gamma_{ik}$ indeed becomes 1 (i.e. $|\gamma_{AB}|=0$, but not all 
$\gamma_{AB}=0$)  at caustics $p=0$. 

The range and the geometrical meaning of the transversal cone coordinates 
$u$ and $v$ must now be discussed. Apparently, $u$ may take all values out 
of the range $(-\infty ,\infty)$. For a real $z$-coordinate the function
$f_2=6 v^2-1-v^4$ cannot have negative values, so $v$ is restricted to an  
interval $(v_{min},v_{max})$ where  $v_{min}=1/(1+\sqrt{2})$ and 
$v_{max}=1+\sqrt{2}$.  If $v$ belongs to this range, also $1/v$ 
does. However, the substitution  (\ref{map}) shows that two different
pairs $(u,v)$ might represent the same null geodesic. Also two different
geodesics might be represented by the same $(u,v)$-pair. 

A natural way to parametrize a light cone is to take the angular 
coordinates $(\theta,\phi)$
at the sky of an observer sitting at the 
vertex and comoving with the fluid. 
As usual, $\theta$ ranges from $0$ (north pole) to $\pi$ (south pole),
and $\phi$ from $0$ to $2 \pi$.                                       
To find the relation between the 
sky coordinates and $(u,v)$ we have used a method,
which is exclusively based on the intrinsic cone metric. Leaving the
details for Appendix B, the result is  
\be
\fl \qquad u^2=\frac{1-\cos{\phi}}{1+\cos{\phi}},\quad  v^2 = 
\frac{\sqrt{2}+\sin{\theta}}{\sqrt{2}-\sin{\theta}},\quad                 
\cos{\phi}=\frac{1-u^2}{1+u^2},\quad \sin{\theta} = 
\sqrt{2}\frac{v^2-1}{v^2+1}.   \label{inverse2}
\ee
$u$ ranges from $u=0$ for $\phi=0$ to $\infty$ at $\phi=\pi$,
jumps there to $-\infty$, and increases to zero at $\phi=2\pi$. $v$
starts from $v=1$ at the north pole ($\theta=0$), increases to 
$v_{max}=1 + \sqrt{2}$ at the equator ($\theta=\pi/2$), and decreases 
to $1$ at the south pole ($\theta=\pi$). As explained in the Appendix,
only the partial interval $(1,v_{max})$ is used for $v$.
A point on the sphere is fixed by a pair $(u,v)$ together with the sign of 
$\epsilon$. 
This ensures that a pair $(u,v)$ from the $u$-range $(-\infty,\infty)$
and $v$-range $(1,v_{max})$ is a one-two-one 
map of the light rays in the northern resp. southern hemisphere. 
The exceptional ray with $v=1, \epsilon=1$ corresponds to the 
north pole, the antipodal ray with $v=1, \epsilon =-1$ to the south pole.

Transforming the inner metric $\gamma_{ik}$ to angular coordinates 
$\theta,\phi$ does not simplify
neither its form nor other relations very much, so we continue 
to work with $u$ and $v$ as transversal coordinates. 

In the chosen representation the equator is  
given by $v=v_{max}=1+\sqrt{2}$ 
and geodesics sent out in these directions (orthogonal to the rotation axis)
lie always in the plane $z=0$. Since here $f_2=0$, the cone metric becomes 
singular,
its determinant $\gamma$ tends to infinity. This is a coordinate singularity: 
The angular coordinates are regular along the equator, but the 
functional determinant $|\frac{\partial (\theta,\phi)}{\partial(v,u)}|$
suffers from a diverging factor $f_2^{-1/2}$. If carefully treated, 
this divergence will not cause trouble.  

\section{Rotation coefficients and invariants}
\subsection{Geometries on null hypersurfaces}

The local differential geometry of null hypersurfaces such as a cone 
was in some detail described in \cite{daut65} and \cite{daut67}, 
see also \cite{pen61},
\cite{pen72}. We summarize the 
points most important for us. This   
geometry is formulated in terms of the rotation 
coefficients of a certain class of triads, defined as follows: At every 
regular point $P$ of the cone there exists a unique direction 
$\epsilon^i,~i=1,2,3$, the direction of the null geodesic passing that point.
$\epsilon^i$ satisfies $\gamma_{ik}\epsilon^k =0$ and is given up to a 
factor by 
$\epsilon^i=\delta^i_1$ in the coordinate system $(u,v,w)$ used in the
last section.
The two other directions, which are spacelike in regular points and 
orthogonal to each other, may be combined linearly to form 
a complex vector $t^i$. We have $t^i\epsilon^k\gamma_{ik}=0$ and 
normalize $t^i$ such that $t^it^k\gamma_{ik}=0,~t^i\bar{t}^k\gamma_{ik}=1$.
The transversal directions $t^i$ are determined only up to a transformation
$t'^i = e^{i\omega}(t^i-\bar{\kappa}\epsilon^i)$, $\omega$ real 
and $\kappa$ complex. 
Note that $\epsilon^i$ is also subject to a change 
$\epsilon'^i=\lambda\epsilon^i$ ($\lambda$ real), since the running 
parameter along a ray may 
be chosen arbitrarily (it need not to be an affine parameter). 
The covariant components of the transversal 
directions are given by $t_i=\gamma_{ik}t^k$ and $\gamma_i$, where 
$\gamma_i$ is defined
by $t^i\gamma_i=0,~\epsilon^i\gamma_i=1$.
This completes the covariant triad.
The rotation coefficients divergence $\rho$, shear $\sigma$ as well
as other coefficients are given in terms of the derivatives of the triad:
\ba
\rho +i\upsilon &=& 
\epsilon^it^k(\bar{t}_{i,k}-\bar{t}_{k,i}),  \label{rdef}  \\ 
\sigma &=&  \epsilon^i\bar{t}^k(\bar{t}_{i,k}-\bar{t}_{k,i}), \label{sdef} \\
\tau &=& \bar{t}^it^k(\bar{t}_{i,k}-\bar{t}_{k,i}), \label{tdef} \\
\chi &=&  \frac{1}{2}\bar{t}^i\epsilon^k(\gamma_{i,k}-\gamma_{k,i}),  \\
i\varphi &=&  \frac{1}{2}\bar{t}^it^k(\gamma_{i,k}-\gamma_{k,i}). 
\ea
$\tau$ is related to the intrinsic geometry of the two-dimensional
wave surfaces $w=const$, with $w$ here as an affine parameter of the 
generating null geodesics. The coefficients $\chi$ and $\varphi$ reflect 
properties of the triad, which are geometrically not relevant. In 
particular, if $\gamma_i$ is chosen as gradient, as we will do here for
simplicity, both $\chi$ and $\varphi$ are zero.
A change of the triad
\ben
t'_i = e^{i\omega}t_i,~ \gamma'_i=\frac{1}{\lambda}\gamma_i
+\kappa t_i + \bar{\kappa}\bar{t}_i,~ 
t'^i = e^{i\omega}(t^i- \bar{\kappa}\lambda\epsilon^i),~
\epsilon'^i = \lambda \epsilon^i  
\een
produces a change of 
the rotation coefficients as follows:
\ba
\rho' &=& \lambda\rho, \nn \\
\sigma' &=& \lambda e^{-2i\omega}\sigma, \nn \\
\tau' &=& e^{-i\omega}(\tau +i\bar{\delta}\omega -i\kappa\lambda D\omega 
-i\kappa\lambda\upsilon +\bar{\kappa}\lambda\sigma -\kappa\lambda\rho), \nn  \\
\upsilon' &=& \lambda(\upsilon + D\omega), \nn \\
\chi' &=& \frac{1}{2}e^{-i\omega}(2\chi+ \frac{\bar{\delta}\lambda}{\lambda}
-\kappa\rho\lambda +i\upsilon\kappa\lambda 
-\bar{\kappa}\lambda\sigma +\lambda D\kappa), \nn \\
i\varphi' &=& \frac{1}{\lambda}i\varphi +\frac{1}{2}(\bar{\kappa}\tau 
-\kappa\bar{\tau}+\delta\kappa -\bar{\delta}\bar{\kappa}) \nn \\
&& +\frac{1}{2}\kappa(2\bar{\chi}+\frac{\delta\lambda}{\lambda} 
 -i\bar{\kappa}\lambda\upsilon -\kappa\lambda\bar{\sigma}
+\lambda D\bar{\kappa}) \nn \\
&& -\frac{1}{2}\bar{\kappa}(2\chi+\frac{\bar{\delta}\lambda}{\lambda} 
 +i\kappa\lambda\upsilon -\bar{\kappa}\lambda\sigma+\lambda D\kappa). \nn
\ea
The last two equations show that  
$\chi=0$ and $\varphi=0$ are preserved for $\kappa,\lambda$ satisfying
$\delta\kappa-\bar{\delta}\bar{\kappa}+\bar{\kappa}\tau - \kappa\bar{\tau}=0,
~\delta\lambda/\lambda^2+D\bar{\kappa}-\bar{\kappa}(\rho+i\nu)
-\kappa\bar{\sigma}=0 $.
Coordinate invariant statements are formulated in terms of those functions of
the rotation coefficients and their derivatives, which are invariant with
respect to the allowed transformations of the triad. The group of 
allowed triad transformations defines the type of null surface 
geometry in the spirit of Felix Klein's "Erlangen program" \cite{klein}.
The most important geometries are the just outlined inner 
geometry and the affine geometry, where the concept 
of an affine parameter for the rays is given as additional geometrical
element.  For other geometries on null hypersurfaces and for more details 
we refer to \cite{daut65} or \cite{daut67}, and for a similar definition of 
null surface geometries the papers by Penrose \cite{pen61}, \cite{pen72} 
should be consulted.

\subsection{Application to the G\"odel cone}

We first determine the rotation coefficients for the G\"odel cone.
The divergence is calculated from 
$\rho =- \frac{1}{4\gamma}\frac{\partial \gamma}{\partial w}$ and may be
written as

\be
\rho = -\cot{2w} 
- \frac{q}{2p\cos{w}},            
\label{rho}
\ee
with the functions $p(v,w), q(v,w)$   defined by (\ref{pfunc}),(\ref{qfunc}).
The divergence tends to $\pm \infty$ at $w=n\pi$ (keels) and $p=0$ (focal
surfaces) and becomes zero at the two-dimensional surfaces 

\be
-\frac{\tan{2w}}{2w} = \frac{f_1^2}{f_2} \label{zerorho}
\ee
between focal surfaces and keels: Since the rhs of this equation is not 
negative,
the range of $w$, where $\rho=0$ is possible, and hence the position of a 
zero-divergence surface,  is restricted by the condition

\be
(2m -1)\frac{\pi}{4} \leq w \leq m\frac{\pi}{2},~~ m=1,2,3... .
\ee
The amount of shear follows most easily from another general 
relation $|\sigma|^2= \rho^2- det(\frac{\partial  
\gamma_{AB}}{\partial w})/(4\gamma)$:  

\be
\fl \qquad |\sigma|^2 = \frac{q^2}{4p^2\cos^2{w}} 
+\frac{8v^2\cos^2{w}+q(2\sin^2{w}-3)}{2p\sin{w}\cos^2{w}} 
+\frac{1}{4\sin^2{w}\cos^2{w}} -\frac{f_1^2}{4v^2}.
\ee
Like the divergence, also the shear goes to infinity at focal surfaces
and keels.
The remaining nonvanishing coefficients may be calculated from 
(\ref{rdef})-(\ref{tdef}), the details are given in Appendix A. 
Splitting $\sigma$ into real and  
imaginary parts, $\sigma=\sigma_1+i\sigma_2,$ one obtains

\ba
\sigma_1 &=&  -\frac{1}{\sin{2w}}+ \frac{f_1^2 \sin{2w}}{2q}
+\frac{q}{2p\cos{w}}, \\
\sigma_2 &=& \frac{f_1^2\sqrt{f_2}\sin^2{w}}{2vq},\\
\nu    &=& -\frac{f_1^2\sqrt{f_2}\sin^2{w}}{2vq}, \\
\tau   &=&  \frac{i\sqrt{f_2}(v^2+1)(2v^2-f_1^2\sin^2{w})}
         {b f_1 p\sqrt{2q}}.
\ea

The triad  has been chosen so that the resulting rotation 
coefficients look as simple as possible.
There exists in general one (and only one) first-order inner invariant of 
a null hypersurface, i.e. an invariant function formed from the rotation 
coefficients alone, without derivatives. This is  
the quantity $j = \frac{\rho}{|\sigma|}$  or any function of $j$. 
It is useful to consider $1/j^2$, which                
measures the anisotropic behaviour of the generators around a given one:
\be
1/j^2 = \frac{pf_1^2(4v^2\sin{w}-p)\sin^2{w}\cos^2{w}+ v^2(p-q\sin{w})^2}
{v^2(p(2\sin^2{w}-1)-q\sin{w})^2}. \label{j2}
\ee

At caustics $p=0$ (and keel points with $w=n\pi$) this 
gives $j^2=1$ or $j= \pm 1$, depending on the sign of $\rho$. Along the 
two exceptional rays (pole rays) the shear vanishes and the anisotropy measure
$1/j^2$ is zero, in accordance with the symmetry properties of the cone.

From the rotation coefficients and their derivatives one may also 
form some second-order invariants.               
The purely transversal  
projections of the four dimensional Ricci and Weyl tensor into the cone
are closely related to them. 
The Ricci and Weyl tensor 
projections are -without any addition of further embedding quantity- 
equal to similar projections related to an  
{\it intrinsic} Riemann tensor $R_{ikl}^{~~m}$
of the null hypersurface,

\ba
t^k\epsilon^l\epsilon^iR_{kli.}^{~~m}\bar{t}_m &\equiv& \omega 
      =D\rho-\rho^2-\sigma\bar{\sigma}, \label{om} \\
\bar{t}^k\epsilon^l\epsilon^iR_{kli.}^{~~m}\bar{t}_m &\equiv& \psi   
      = D\sigma -2\sigma(\rho -i\upsilon). \label{psi}
\ea
A straightforward calculation gives
\ba
\omega &=& \frac{(v^2+1)^2}{4v^2}, \label{omex} \\
\psi & = & f_1^2(\frac{4\cos^2{w}}{q}- \frac{1}{2v^2}
+ 2i\frac{\sqrt{f_2}\sin{w}\cos{w}}{vq}). \label{psiex}
\ea

To calculate an intrinsic curvature tensor of a null hypersurface as in 
(\ref{om},\ref{psi}), one needs an affine connexion in spite of the 
missing unique contravariant metric tensor. We again refer to 
\cite{daut65},\cite{daut67} and note only shortly, 
that in general the resulting affine connexion and hence the inner 
Riemann tensor depend on the triad. Independence holds only for 
special projections such as (\ref{om},\ref{psi}).
$\omega$, $\psi$ are nevertheless not yet affine or inner {\em invariants}
of the cone, they are {\em densities}, and one has to 
apply suitable factors of $\rho$ or $|\sigma|$ to generate invariants of 
the {\it affine} geometry. To obtain invariants of the {\it inner} geometry, 
one has to take a certain linear combination of these affine invariants.   
If we define
\be
\fl \qquad \qquad I=I_1+iI_2 = i(\frac{\omega}{\rho|\sigma|}
-\frac{\psi}{\sigma|\sigma|} +
\frac{1}{j}-j)=\frac{i}{|\sigma|}\left (\frac{D\rho}{\rho}
-\frac{D\sigma}{\sigma}\right )+\frac{2\nu}{|\sigma|}, 
\ee
then this quantity {\it is} a second-order differential invariant
of the inner geometry. 
A short calculation gives for the real part 
$I_1=(Ds+2\nu)/|\sigma|$
(where $s$ is the argument of $\sigma$ such that 
$\sigma=|\sigma| e^{is}$)
\be
|\sigma|^3 I_1  = \frac{f^2_1\sqrt{f_2}(4v^2\sin{w}-p)}{4pv^3}. 
\ee
$I_1$ is a measure of the rotation of the two shear directions 
(defined as directions to neighbour rays with extremal distance change, 
\cite{daut65}) with regard to the generator congruence.
The imaginary part, which can also be written $I_2= Dj/\rho$, 
is slightly more complicated and may be 
represented as  
\be
|\sigma|^3 I_2 = \frac{i_0+i_1 p+i_2p^2+i_3p^3}
{4p^2 v^2 \sin^2{w}\cos{w} (q\sin{w} +p(1-2\sin^2{w}))} 
\ee
with 
\ba
i_0 &=& 4v^2q^2 (q-2v^2)\sin^2{w}, \nn \\
i_1 &=&f_1^4(v^2+1)^2\sin^7{w}-8v^2f_1^2(v^2-3)(3v^2-1)\sin^5{w} \nn \\
& & +64v^4(f_1^2-v^2)\sin^3{w}-64v^6\sin{w},     \\
i_2 &=& -2f_1^2f_2\sin^4{w}+12v^2f_2\sin^2{w}+ 8v^4(1-2\sin^2{w}),\nn  \\
i_3 &=& -f_2\sin{w}. \nn
\ea

The Gaussian curvature 
$K = -2\tau\bar{\tau} +\delta\tau+\bar{\delta}\bar{\tau}$ (see \cite{daut65}) 
of the two-dimensional surfaces $w=const$ 
is in general not an invariant of null hypersurfaces. The only 
exception are Killing horizons (sometimes called "totally geodesic null 
hypersurfaces", see, e.g.,  Hajicek \cite{hajicek}), defined by the condition 
that the inner metric admits a 
Killing symmetry with the generators as Killing vectors. In all other
cases $K$ depends on the chosen foliation and is 
not significant for the cone geometry: A change of the affine parameter as
$\bar{w}=a(u,v)w$ (keeping the vertex at $\bar{w}=0$) leads to a different
curvature $\bar{K}$.    
$K$ is only invariant under transformations of the 
 transversal parameters $x^A{'}=f^A(x^B)$.
For our foliation $w=const$ an explicit calculation gives
\be
K= \frac{k_1}{p^2} +\frac{k_2}{p^3} \label{gcurv}
\ee
with 
\ba
k_1 &=& (32v^6-f_1^6\sin^2{w})/(b^2f_1^2),\\
k_2 & =&  4f_2v^2(v^2+1)^2\sin{w}(\sin^2{w}f^2_1-2v^2)/ (b^2f_1^2).  
\ea
$K$ generally tends to zero for large $w$, apart 
from spikes at focal surfaces $p=0$. 

In \cite{daut65},\cite{daut67} points on a given ray have been 
classified according to the focussing 
behaviour of neighbouring geodesics. A point on a ray was called 
elliptic, if the spatial distance to {\it all} neighbouring rays either 
increases or decreases, and hyperbolic, if some
rays converge and other diverge. The sign of $\rho^2-|\sigma|^2$ distinguishes 
both types of points. A positive sign (or $j^2>1$ at points with nonvanishing 
shear $|\sigma|$) corresponds to elliptic points.
Evidently, near the vertex at $w=0$ all points on all rays 
are elliptic, this is also seen from an expansion of $j$ near the vertex, 
$j \approx -6v^2/(f_1^2w^2) $.
Zeros or infinities of $\rho^2-|\sigma|^2$ 
 along a ray may  signify the transition 
from elliptic to hyperbolic points (or vice versa, a point with 
$\rho^2-|\sigma|^2=0$ will be called parabolic).
It is not difficult to verify that $\rho^2-|\sigma|^2$ can be written as 
\be
\rho^2-|\sigma|^2 = \frac{2(2v^2-f_1^2\sin^2{w})}{p\sin{w}}-
  \frac{f_2}{4v^2}.
\ee
Thus caustic singularities ($p=0$ or $w=n\pi$)
can be transition points on a ray. We illustrate this for the two
pole rays ($v=1$) and for the equator rays ($v=1+\sqrt{2}$). For 
pole rays  $\rho^2-|\sigma|^2= \cot^2{w}$,  all ray points are elliptic, 
with the exception of isolated parabolic points at $w= n\pi +\pi/2$,  
hence no proper transition point exists.
In the case of equator rays we have $\rho^2-|\sigma|^2= 2\cot{2w}/w$,  
here all points with $w=\pi(2n-1)/4$, $n$ integer, are transition points,
including also non-caustic points.

To have some idea of the cone structure, we follow a typical 
ray with $1<v<1+\sqrt{2},~u$ arbitrary, from the vertex
$w=0$ down the cone (we consider the past light cone, where an
increasing affine parameter $w$ means decreasing time).
Near the vertex, the cone resembles the Minkowski light cone with
$\rho=-1/w$, zero shear and exclusively elliptic points. All neighbour 
rays recede from the chosen ray. Along the ray, $|\rho|$ decreases from 
$\rho =-\infty$ at $w=0$  and $|\sigma|$ increases from zero, until a 
first transition point is reached, where both are equal. At the transition 
point the invariant $j$ has increased from $-\infty$ at $w=0$ to $j=-1$. 
Behind this point a domain of hyperbolic points begins, where some neighbour 
rays start to decrease their distance to the chosen ray.
The next significant point on the ray is the zero-divergence point, where also
$j$ reaches zero for the first time. After passing this point, $\rho$ is
positive and increases faster than $|\sigma|$. Thus a second transition point 
is encountered with $j=1$, ending the domain of hyperbolic points. Behind the 
transition point a region of elliptic points begins, all rays converge towards
our ray, 
preparing for a meeting at the first focal point. 
At the focal point both $\rho$ and $|\sigma|$ tend to infinity, their quotient
$j$ jumps from $1$ to $-1$.
 
Behind the focal point, $\rho$ increases from large negative values, passing 
a third  zero-divergence
point in the interval $ 3\pi/4<w<\pi$, until the first keel point is
reached at $w= \pi$. The whole region between focal surface and 
keel consists again of hyperbolic points. 
Behind the keel point the cycle starts again, with changed 
positions of the transition and focal points relative to the zero-divergence 
and keel points.

\section{Focal sets}                    
\subsection{Keels}

The most impressive  singularities of the G\"odel cone are those at 
$w=n\pi$ (Fig.2), the keel points. They correspond to "points of the first 
kind" on the light cone of \oss anti-Mach-metric \cite{os62}. 
At the keel point all rays with equal $v$ and different $u$ meet . 
An one-dimensional set of 
connected keel points is denoted as keel line.  Every integer $n$ gives a 
keel line with the spacetime coordinates 
\be \fl \qquad \quad
t_{keel}= \frac{n\pi b(v^2+1)}{\sqrt{2} v},~         
x_{keel}= 0,~       
y_{keel}= b,~        
z_{keel}= -\frac{n\pi b\epsilon \sqrt{6 v^2-1-v^4}}{\sqrt{2} v}.   
\ee 
Thus keel lines can be considered as 
circle segments in the pseudo-Euklidean 
$z-t$ plane with a length  $L$ increasing with $n$ (Fig.3).

\begin{figure}[tbp]
\includegraphics[width=5.5cm]{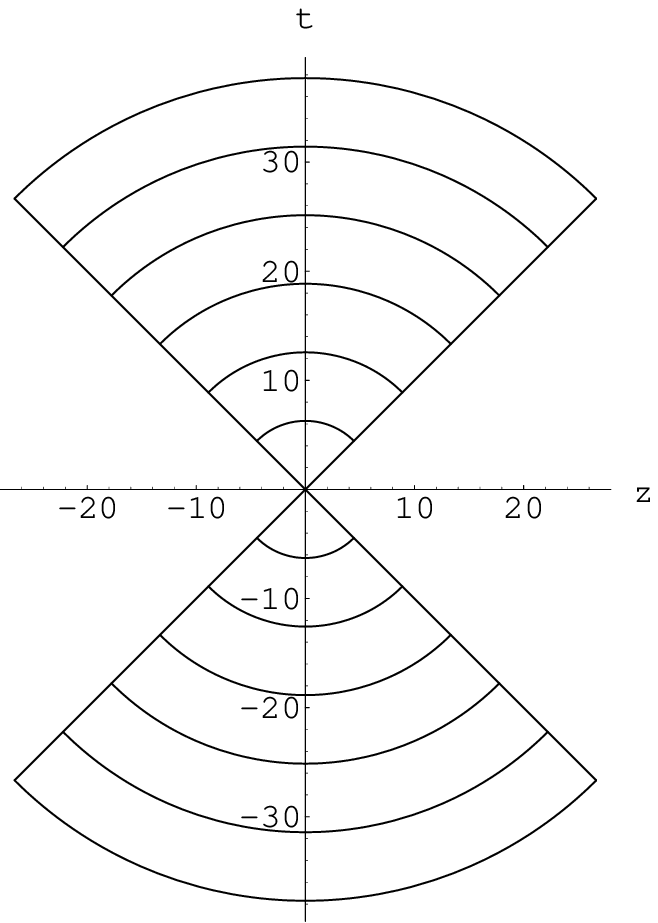}
\caption[fig3]\
\small 
\baselineskip=10pt
\smallskip
{The keel lines $w=n\pi$ appear as sequence of isolated sphere segments 
with increasing length, if projected into the pseudo-Euklidean plane 
$z-t$.  Shown are the keel segments for the future and past cone 
($n=1...6$, $b=1$). The $t$-axis is the projection of the observer world
line. The endpoints of the keel lines lie on the 
exceptional ray in the northern hemisphere ($z>0$) and on its 
antipodal ray in the southern hemisphere ($z<0$). 
These rays are plotted as straight lines.
The keel endpoints 
are also the intersection points of the $n$th keel line with   
the $n$th focal surface. 
Every point on a keel line     
(which is parametrized by $v$) is an intersection point of all 
generators with different parameters $u$ (and the same $v$). In 
particular, the intersection of the keel with the $t$-axis corresponds
to equator    rays $v=v_{max}$, the only rays which return to the observer 
worldline.}

\hrulefill
\end{figure}

Their end points lie on the exceptional ray and its antipode.  
For $n=0$ the keel lines shrink to the vertex $P_0$, and for $n \neq 0$ the 
observer world line crosses the keel lines only for equator rays 
 $v=v_{max}$.  
It is easily checked that the keel 
lines are spacelike but not
geodesic in the G\"odel geometry. The components of the tangential vector 
$dx^{\mu}/dv$ are
\be
\left(\frac{        dx^\mu}{        dv}\right)_{keel} 
=\frac{n\pi b}{v^2\sqrt{2}}\left(1-v^2,0,0,\frac{\epsilon(v^4-1)}
{\sqrt{6v^2-v^4-1}}\right),
\ee
and its norm is $b^2n^2\pi^2 f_1^4/(v^4 f_2)$. 
The first normal of the keel (the binormal does not exist) 
is the timelike unit vector
\be
n^\mu_1 = \frac{1}{\sqrt{2} (v^2-1)}(v^2+1, 0,0,\epsilon\sqrt{6v^2-1-v^4}),
\ee
the (first) curvature is found as 
\be
k_1 = \frac{4v^4}{nb\pi (v^2-1)^3},
\ee
it diverges at the
two endpoints $v=1$ of the keel lines. Here also 
the otherwise spacelike tangential vector degenerates to zero.
The invariant total 
length of a keel segment, ranging over the full observer sphere, is
however finite and given by
\be
L = \int ds = 2n\pi b\int_{1}^{1
+\sqrt{2}}\frac{(v^2-1)^2}{v^2\sqrt{6v^2-v^4-1}}\,dv \approx 2.39628~n \pi b.
\ee
The divergence of the integrand at the equator $v=1+\sqrt{2}$ 
arises from the coordinate singularity there,  
but the integral converges (we could equally well have integrated from
$v=1/(1+\sqrt{2})$ to $v=1$ with the same result).

It is of interest to study the behaviour of rotation 
coefficients and in particular differential 
invariants near the 
singularities  $p=0$ and $ w=n\pi$. 
A power series expansion 
around keel points $w=n\pi$, $n \neq 0$, leads to              
\ba
\rho &=& -\frac{1}{2(w-n\pi)}- \frac{2v^2}{n\pi f_1^2} + o((w-n\pi)), \\
|\sigma| &=& \frac{1}{2|w-n\pi|}
- \frac{2v^2}{n\pi f_1^2} U(w-n\pi) + o(w-n\pi), \\
j &=& -U(w-n\pi)   + \frac{8v^2}{n\pi f_1^2}(w-n\pi))+ o((w-n\pi)^2) 
\ea
for small $w-n\pi$, where $U(x)$ is the step function, $U(x)=1$ for $x>0$ 
and $-1$ for $x<0$. While the ray divergence as well  as the shear 
amount individually have first order poles
at focal or keel points, their invariant quotient $j$ remains finite, but jumps
from $+1$ to $-1$, if $w$ increases.         
For the second order invariants $I_1,I_2$ one obtains near keel 
points: 
\ba 
I_1 &=&-\frac{2f_1^2\sqrt{f_2}}{v^3}(-1)^n(w-n\pi)^3 + o((w-n\pi))^4, \\   
I_2 &=& \frac{16 v^2}{n\pi f_1^2}(-1)^n(w-n\pi) + o((w-n\pi)^2),
\ea
thus the complex invariant $I=I_1+iI_2$ {\it vanishes} here.

\subsection{Focal surfaces}

As already discussed in previous sections, apart from the keel lines   
also focal singularities exist       
(we shall not discuss coordinate singularities).   
In geometrical optics a caustic (set of focal points) is the locus where the 
rays have an envelope and the intensity a singularity. Here 
focal point are similiarly defined as points of intersection of 
infinitesimally close geodesics, which satisfy the relations 
$x^\mu(u+\delta u,v+\delta v,w+\delta w) -x^\mu(u,v,w)=0$, where 
$x^\mu(u,v,w)$ is the cone congruence (\ref{tc})-(\ref{zc}). 
Expanding, we have  
\be
\frac{\partial x^\mu}{\partial u}\delta u+
\frac{\partial x^\mu}{\partial v}\delta v+
\frac{\partial x^\mu}{\partial w}\delta w =0 \label{fe} 
\ee
for suitable displacements $\delta u, \delta v, \delta w$.
Let us assume $v \neq 1$ (we let out the exceptional rays) and
$\sin{w}\neq 0$ (no keel points are considered). We can then 
eliminate the displacements from (\ref{fe}) and obtain two relations, 
which must be 
satisfied for the coordinates $u,v,w$ of focal points on the cone:
\ba
p(v,w)r(u,v,w) &=& 0,\\
p(v,w)s(u,v,w) &=& 0.
\ea
Here $r=\sum{r_iu^i}$ and $s=\sum{s_iu^i}$  are polynomials of 
fourth order in $u$, $r_i$ and $s_i$ are polynomials in
$v,w,\sin{w}$ and $\cos{w}$. A closer inspection shows that with the
restrictions $\sin{w} \neq 0, v \neq 1$, $r$ and $s$ cannot vanish 
simultaneously, thus we conclude that focal points are given by 
$p(v,w)=0$ or
\ben              
(v^2-1)^2w\cos{w}+(6v^2-1-v^4)\sin{w}=0,            
\een
which might also be written as 
\be
-\frac{\tan{w}}{w} = \frac{f_1^2}{f_2} = \frac{1}{2}\tan^2{\theta}. 
\label{fp}
\ee
Alternatively, focal points can be considered as the critical
points of the map $(u,v,w) \rightarrow x^\mu$, i.e, points
where the rank of the Jacobian matrix is not maximal \cite{arnold85}.  
This leads to the same condition (\ref{fp}).
(\ref{fp}) is the equation of two-dimensional surfaces on the cone. It
is similar to the condition for "points of the second kind" on the 
Ozsv\'{a}th-Sch\"ucking cone \cite{os62}. Note also the similarity of this
equation to the equation for zero-divergence surfaces (\ref{zerorho}).
As in this case we conclude that focal surfaces can only occur in 
regions, where the affine parameter $w$ is confined to the intervals
\be
(2 n-1)\frac{\pi}{2} \leq w \leq n \pi,~~ n=1,2.3..., \label{finterv}
\ee
since only here $-\tan{w}/w$ is not negative.
Thus the focal set decays into an infinite number of separated 
two-dimensional sheets. 
The circular functions generate a quasi-periodic 
behaviour with similar but not identical shapes for the sheets.         
 
If we solve (\ref{fp}) for $v$,
we find
\be
v^2 = 1 +\frac{2(\sin{w} +(-1)^{n+1}\sqrt{2\sin^2{w}-
w\sin{w}\cos{w}})}{\sin{w}-w\cos{w}}.
\label{piece}
\ee
If $w$ is in the interval (\ref{finterv}), the square root is real.
Since $v^2-1$ varies only between $0$ and $2+2\sqrt{2}\approx 4.8284$,
the quotient in (\ref{piece}) must fit into this interval. 
This is achieved by choosing the sign of the square root as indicated.

Keels and focal surfaces are not completely separated. 
Every keel line $w=n\pi$ 
has two common points with the $nth$ sheet of 
focal surfaces, corresponding to the keel line endpoints $w=n\pi, v=1$. 
This common point lies on the exceptional null ray resp. its antipode,  
see also Figure 3. 

If (\ref{piece}) is inserted into the equations (\ref{tc})-(\ref{zc}), 
one obtains an  explicit representation 
$x^\mu_{focal}(u,w) =x^\mu(u,v[w],w)$ of the focal surfaces. The tangential
directions at the point $(u,w)$ on a focal surface
are spanned by the spacelike vector 
$\partial x^{\mu}_{focal}/\partial u$ and the vector   
\be
k^\mu = \frac{\partial x^\mu}{\partial w} + 
\frac{dv^*}{dw} \frac{\partial x^\mu}{\partial v} + 
\frac{dv^*}{dw}
\frac{\partial x^\mu}{\partial u} \frac{(u^2+1)}{f_1\cos{w}},
\ee
where $v^*(w)$ is the function defined by (\ref{piece}).
$k^\mu$ is in general spacelike on the cone, but  becomes a null vector at  
focal surfaces. We expect that the curves 
to which $k^\mu$ is tangent are non-geodesic null curves (see 
\cite{bonnor69} for an excellent discussion of non-geodesic null lines in the
Minkowski spacetime). 

Considering the invariants near focal surfaces, it turns out that $j$ 
jumps from $+1$ to $-1$ with increasing $w$ - this is the same behaviour 
as in keel points.
Near the $n$th focal surface $p=0$ we may write with the step function $U(p)$:
\be
  j = (-1)^nU(p)(1 -\frac{4\cos^2{w}(f^2_1\sin^2{w}-2v^2)}
{q^2\sin{w}}p) +o(p^2). 
\ee
Note that with increasing $w$ the function $p(v,w)$ reaches zero at the 
$n$th focal surface from positive (negative) values, if $n$ is odd (even).
A similar calculation for the invariants $I_1,I_2$ gives 
\ba 
I_1= \frac{8(-1)^{n+1}f_1^2\sqrt{f_2}\cos^3{w}\sin{w}}{vq^3}p^2+o(p^3),\\   
I_2 = \frac{8\cos^2{w}(2v^2 - f_1^2\sin^2{w})}{q^2\sin{w}}|p|
+ o(p^2)
\ea
near the focal surfaces $p=0$. 

This shows that at both focal and keel points the complex invariant 
$I= I_1+iI_2$ vanishes, while $j$ is $1$, modulo a sign. 
The same result holds for the  
 Oszv\'{a}th-Sch\"ucking lightcone \cite{os62}, see \cite{abdel72a}.

\ack                         
We are indepted to J. Ehlers for discussions. 
One of the authors (A.-M.) is grateful to the Albert Einstein Institute 
for hospitality during the preparation of the paper.

\appendix

\section{A triad on the G\"odel cone}

For an explicit calculation of the rotation coefficients (\ref{rdef}),
(\ref{sdef}), (\ref{tdef}) one needs the components of a suitable triad
$\epsilon^k,t^k,\bar{t}^k$ and $\gamma_k,t_k,\bar{t}_k$ on the light 
cone. We have already chosen $\epsilon^i=\delta^i_1$.  The inner metric
is given in terms of the triad by
\be
\gamma_{ik}= t_i\bar{t}_k + \bar{t}_it_k 
\ee
Comparing this expression with (\ref{im000})-(\ref{im033}) shows that
$t_i=\delta^A_it_A,~ t^i=\delta^i_A t^A,~ A=2,3 $. It is not difficult to  
verify that 

\ba
t_2 &=& \frac{b\sin{w}(p-q\sin{w})}{v^2\cos{w}\sqrt{2q}} 
+ i\frac{bp}{v} \sqrt{\frac{2}{qf_2}},   \\
t_3 &=& \frac{\sqrt{q}b f_1 \sin{w}}{\sqrt{2}v^2(u^2+1)}
\ea
reproduces the equations (\ref{im22})-(\ref{im33}). For the contravariant
components we use the normalization conditions $t^it^k\gamma_{ik}=0,~
t^i\bar{t}^k\gamma_{ik}=1$ and obtain
\ba
t^2 &=& i\frac{v}{2bp}\sqrt{\frac{f_2q}{2}}, \\                         
t^3 &=& \frac{v^2(u^2+1)}{bf_1\sin{w}\sqrt{2q}} 
+i\frac{v(u^2+1)\sqrt{f_2}(q\sin{w}-p)}{2bf_1\cos{w}\sqrt{2q}p}.                                                     
\ea
The nonvanishing rotation coefficients may then be found from
\ba
\rho + i\nu &=& -t^2\bar{t}_{2,1}- t^3\bar{t}_{3,1},\\
\sigma &=&  -\bar{t}^2\bar{t}_{2,1}- \bar{t}^3\bar{t}_{3,1},\\
\tau &=&  (\bar{t}^2t^3-\bar{t}^3t^2)(\bar{t}_{2,3}- \bar{t}_{3,2}).  
\ea

\section{Observer sky and cone parametrization}

In sky coordinates $\theta,\phi$ any cone metric can be expanded in
powers of an affine parameter $w^*$  near the vertex
(see, e.g., \cite{daut65}):
\ba
\gamma_{\theta\theta}^* &=& \frac{w^{*2}}{2} + o(w^{*4}), \nn \\
\gamma_{\theta\phi}^* &=&  o(w^{*5}), \nn \\
\gamma_{\phi\phi}^* &=& \frac{w^{*2}}{2}\sin^2{\theta}   + o(w^{*4}). \nn
\ea
A similar expansion of the G\"odel cone metric in powers of $w$ gives
\ba
\gamma_{22} &= & \frac{4w^{2}b^2f_1^2}{v^2(u^2+1)^2} + o(w^{4}), \nn \\
\gamma_{23} &= &  o(w^{5}), \nn \\
\gamma_{33}  &= & \frac{16w^{2}b^2}{f_2} + o(w^{4}).\nn
\ea
The coordinates $(u,v,w)$ are related to $(\theta,\phi,w^*)$,
and this coordinate transformation should take
approximately the form $\theta=\theta(u,v),~\phi=\phi(u,v),
~w^*=w/m(u,v)$ near the vertex, i.e., for small $w$.
For the transformation functions we thus obtain
the differential equations
\ba
\left(\frac{\partial\theta}{\partial u}\right)^2+
\left(\frac{\partial\phi}{\partial u}\right)^2 \sin^2{\theta} &=&
\frac{8b^2f_1^2m^2}{v^2(u^2+1)^2}, \label{dg1} \\
\frac{\partial\theta}{\partial u}\frac{\partial\theta}{\partial v}+
\frac{\partial\phi}{\partial u}\frac{\partial\phi}{\partial v}\sin^2{\theta}
&=&0, \label{dg2} \\
\left(\frac{\partial\theta}{\partial v}\right)^2 +
\left(\frac{\partial\phi}{\partial v}\right)^2
 &=& \frac{32b^2m^2}{f_2}, \label{dg3}
\ea
which can easily be solved, if we assume
$\theta=\theta(v), \phi=\phi(u)$. Then (\ref{dg2}) is already
satisfied and the other two give
\ben
(u^2+1)\frac{\partial\phi}{\partial u}  =\frac{2\sqrt{2} f_1bm}{v\sin
{\theta}}, \qquad \frac{\partial \theta}{\partial v}=
 \frac{4\sqrt{2}\epsilon_1bm}{\sqrt{f_2}},
\een
where $\epsilon_1 =\pm 1$.
The second equation here shows that $m$ depends on $v$ only, the first
equation then says that both sides must be equal to a constant $k_1$
independent of $u$ and $v$. Integrating, we first obtain
\ben
u = \tan{(\phi/k_1)}.\nn
\een
The differential equation for $\theta$ follows
as
\be
\frac{1}{\sin{\theta}}\frac{\partial\theta}{\partial v}
= \frac{2\epsilon_1k_1v}{f_1\sqrt{f_2}} \label{pheq}
\ee
and is solved by
\be
\tan{\frac{\theta}{2}}= k_2\left(\frac{f_1}{1+v^2
+\sqrt{f_2}}\right)^{\epsilon_1k_1/2}, \label{theq}
\ee
where $k_2$ is a second constant. To obtain a real
square root of $f_2$, $v$ had to be confined to the interval
$v_{min}=1/(1+\sqrt{2})$ through $v_{max}=1+\sqrt{2}$.
We refer to $(v_{min},1)$ as the {\it min} interval
and to $(1,v_{max})$ as the {\it max} interval. 

To determine the coefficients $k_1,k_2$ in 
Eq. (\ref{theq}), we must face the possibility 
that they differ in different parts of the sphere.
 
The symmetry properties suggest that the poles of the observer 
sphere are related to the local rotation axis and are thus 
given by $v=1$.
We consider (\ref{theq}) near the north pole and 
assume, that the $v$ 
belong to the {\it max} range.
Expanding the rhs in powers of small $v-1$, on obtains 
$k_2((v-1)/2)^{\epsilon_1k_1/2}$, thus 
the sign of $\epsilon_1 k_1$ must be positive to
ensure that the rhs vanishes for $v\rightarrow 1$, as does the lhs.
We also conclude that $k_2>0$.
A similar conclusion is reached for the {\it min} range of $v$:
Apart from the positive sign of $\epsilon_1 k_1$, also 
$k_2(-1)^{\epsilon_1k_1/2}$ must be positive and real.  
(Note, the {\it min}-interval of $v$ is obtained from the 
{\it max}-interval by applying 
the map $v \rightarrow 1/v$, the expression within the bracket in 
(\ref{theq}) attains the factor -1 under this map).
Using the identity 
$\tan{(\frac{\theta}{2})}\tan{(\frac{\pi}{2}-\frac{\theta}{2})} =1$,
one can easily repeat the calculation near the south pole.        
It is seen that the lhs of (\ref{theq}) diverges 
there, thus 
also the rhs diverges, and this requires $\epsilon_1k_1 < 0$, 
holding again for {\it min} as well as for {\it max} ranges of $v$. 
Furthermore, we have for the {\it max} ({\it min}) 
range  $k_2>0$ ($k_2(-1)^{-k_3/2}>0$).

Moving now from the north pole towards the equator, assuming
the {\it max} interval, $v$ as well as $\theta$ increase   
until $v=v_{max}$ is reached, which  corresponds  
to a $\theta_{max}= 2\arctan{[k_2/2^{\epsilon_1k_1/2}]}$.
$\theta_{max}$ cannot represent the other pole 
$\theta=\pi$, since 
the lhs of (\ref{theq}) 
diverges at $\theta= \pi$, while the rhs here is regular.
Thus only part of the sky is covered by $v$ values in the 
{\it max} range. It is convenient to assume that this part is the 
northern hemisphere, i.e., $\theta_{max}=\pi/2$. This fixes 
$k_2$ by $k_2= 2^{\epsilon_1k_1/4}$. If we had started our walk 
in the {\it min} region of $v$, taking $v=1$ at the north pole and 
decreasing $v$ to the equator,    we would have obtained a similar
conclusion: Also the {\it min} range covers the sphere only from the 
pole to the equator, here $k_2$ is fixed by 
$k_2=2^{\epsilon_1k_1/4}(-1)^{-\epsilon_1k_1/2}$.  
Our walk could have started from the south pole, reaching 
the equator from the south, but 
the results for $k_2$ are the same.
 
Further conclusions depend on $k_1$. Mapping the $(0,2\pi)$ interval of 
$\phi$ to the range $(0,\infty)$ of $u$ would mean $k_1=4$, but this cannot 
be correct: Since the meridians $\phi=0$ and $\phi=2\pi$ and hence  $u=0$ 
and $u=\infty$  coincide, the corresponding rays must represent 
the same spacetime points, which is wrong, as a discussion of 
(\ref{tc})-(\ref{zc}) shows. The correct choice is $k_1=2$, which  maps
$(0,2\pi)$ to the $u$-interval $(-\infty,\infty)$ in the sense that
$(0,\pi)$ is mapped to $(0,\infty)$ and $(\pi,2\pi)$ to 
$(-\infty,0)$. Since the subsets  $(v,u\rightarrow\infty)$
and   $(v,u\rightarrow -\infty)$ in the parameter space of $u$ and $v$ 
describe the same rays, there is no matching problem here.      
Taking only {\it max} regions for the 
$v$-values and assuming  $\epsilon_1=1,~k_2=\sqrt{2}$
for the northern and $\epsilon_1=-1,~k_2=1/\sqrt{2}$ for the 
southern hemisphere will satisfy our conditions.                  
Note that the northern and southern hemisphere need separate copies of the
{\it max} interval.
$u$ covers the range $(-\infty,\infty)$,
and the sign $\epsilon_1$ turns out to be equal to $\epsilon$
in Eq. (\ref{zc}).
Thus the pair $(\theta,\phi)$ is related to a pair $(u,v)$ by
\be
\tan{\frac{\theta}{2}}  =  \left(\frac{\sqrt{2}(v^2-1)}{v^2 +1
+ \sqrt{6v^2-v^4-1}}
\right)^\epsilon, \qquad
\tan{\frac{\phi}{2}}  =   u, \label{ph}
\ee
where
$\epsilon= 1(-1)$ in the northern (southern) hemisphere. 
 (\ref{ph}) may be inverted. Thus finally we have
\be  
v^2 = \frac{\sqrt{2}+\sin{\theta}}{\sqrt{2}-\sin{\theta}},\qquad 
u^2=\frac{1-\cos{\phi}}{1+\cos{\phi}}, \label{inverse}
\ee 
valid for the {\it whole} sphere.
We can completely discard the {\it min} regions. 
Rays with 
$v$ from the {\it min} region are also light cone rays, 
but the transformation $v \rightarrow 1/v, u \rightarrow -1/u$ 
leads to identical rays, compare Eq.(\ref{map}), 
thus already all rays are covered, if we confine the discussion to 
{\it max} intervals. It should nevertheless be noted that another   
parametrization is possible, which makes use of {\it min}-intervals.



\section*{References}

\begin{thebibliography}{99}

\bibitem{abdel72a}    Abdel-Megied M 1972
 \"Uber die Lichtkegel in speziellen kosmologischen Modellen, 
PhD thesis, Humboldt University at Berlin        

\bibitem{abdel72b} Abdel-Megied M  and Dautcourt G 1972      
Zur Struktur des Lichtkegels im G\"odel-Kosmos  
{\it  Mathematische Nachrichten} {\bf 54}, 33--39         

\bibitem{arnold85} Arnold V I, Gusein-Zade S M and Varchenko A N
1985 {\it Singularities of Differentiable Maps}  Vol. I   
(Boston, Basel, Stuttgart: Birkh\"auser)      

\bibitem{arnold90} Arnold V I 1990  {\it Singularities of Caustics and 
             Wavefronts}  (Dordrecht: Kluwer)   

\bibitem{bonnor69} Bonnor W B 1969  Null Curves in a Minkowski Space-Time,
{\it Tensor} {\bf 20}  229--42          

\bibitem{daut65} Dautcourt G 1965
{\it Nullfl\"achen in der allgemeinen 
Relativit\"atstheorie}, 
habilitation thesis, Humboldt University at Berlin            

\bibitem{daut67} Dautcourt G 1967
Characteristic hypersurfaces in general relativity  
{\it J. Math. Phys.} {\bf 8}  1492--501              

\bibitem{cork83} Corkill R W  and Stewart J M 1983
Numerical Relativity. II. Numerical methods for the characteristic 
initial value problem and the evolution of the vacuum field equations
for space-times with two Killing vectors 
{\it Proc. R. Soc. London} {\bf A 386} 373--91 

\bibitem{ehl00}
Ehlers J and Newman E T 2000      
The theory of caustics and wave front singularities with physical applictions
{\it J. Math. Phys.} {\bf 41} 3344--78        

\bibitem{friedr83} Friedrich H and Stewart J 1983 
Characteristic initial data and wave front singularities in general relativity
{\it Proc. R. Soc. London} {\bf A 385} 345--71 

\bibitem{fritt03}        
Frittelli S and Petters A O 2003 Wavefronts, Caustic Sheets, and 
Caustic Surfing in Gravitational Lensing
{\it J. Math. Phys.} {\bf 43}  5578--611

\bibitem{godel49}
G\"{o}del K 1949  An example of a new type of cosmological solution 
of Einstein's field equations of gravitation 
{\it Rev Mod Phys} {\bf 21}  447-50

\bibitem{godel52}
G\"odel K 1952 Rotating universes 
{\it Proc Int Cong Math}(Camb, Mass). Ed. L.~M.~Graves et al. 
{\bf 1}, 175 

\bibitem{hajicek} Hajicek P  1973  
Exact models of charged black holes: 
I. Geometry of totally geodesic null hypersurface.   
{\it Comm. Math. Phys.} {\bf 34}  37--52

\bibitem{hawell73}
Hawking S W and Ellis G F R  1973   
{\it The Large Scale  Structure of Space-Time} (Cambridge: Cambridge 
University Press)

\bibitem{klein} Klein F and Wussing H 1974 {\it Das Erlanger Programm}
(Leipzig: Teubner)

\bibitem{kundt56}
Kundt W 1956 Tr\"agheitsbahnen in einem von G\"odel angegebenen  
kosmologischen Modell {\it Zs f Ap} {\bf 145} 611-20

\bibitem{laur81} Laurent B E, Rosquist K  and Sviestins E  1981   
The Behaviour of Null Geodesics in a Class of Rotating Space-Time
Homogeneous Cosmologies {\it Gen.Rel.Grav.}{\bf 13} 1093

\bibitem{os62}
Oszv\'{a}th I and Sch\"{u}cking E  1962   An anti-Mach Metric, in
{\it Recent Developments in General Relativity} (New York: Pergamon Press) 
339--50

\bibitem{os70}
Oszvath I 1970 Dust-Filled Universes of Class II and Class III
{\it J. Math. Phys.} {\bf 11 }  2871-83

\bibitem{os03} Oszv\'{a}th I and Sch\"{u}cking E 2003 G\"odels trip
{\it Am. J. Phys.} {\bf 71} 801-05 

\bibitem{pen65} Penrose R 1965  A remarkable property of plane waves in
general relativity {\it Rev. Mod. Phys.} {\bf 37} 215 

\bibitem{pen72} Penrose R 1972
        The geometry of impulsive gravitational waves 1972
        {\it General Relativity, Papers in Honour of J.\,L. Synge,}
        edited by L. O'Raifeartaigh (Oxford: Clarendon Press)
        p~101-15

\bibitem{pen61} Penrose R 1961 Null Hypersurface Initial Data for
Classical Fields of Arbitrary Spin and for General Relativity,
published as Golden Oldie {\it Gen. Rel. Grav.} {\bf 12} 225

\bibitem{perlick04}
Perlick V  2004  Gravitational Lensing from a Spacetime Perspective
{\it Liv. Rev. Rel.} 2004-9                                   

\bibitem{riesz} Riesz M 1956
Problems related to characteristic surfaces     
{\it Proc. Internat. Conf. in Differential Equations} {\bf 57}            

\bibitem{rooman98} Rooman M and Spindel Ph 1998  G\"odel metric as a 
squashed anti-de~Sitter geometry
    {\it Class Quantum Grav} {\bf 15} 3241--49

\bibitem{schn92} Schneider P, Ehlers J and Falco E E 1992  Gravitational 
Lenses (New York, Berlin, Heidelberg: Springer-Verlag)

\bibitem{winic98}
Winicour J 2001 Characteristic Evolution and Matching    
{\it Liv. Rev. Rel.} 2001-3                                  

\end{thebibliography}
\end{document}